\documentclass[11pt,a4paper]{article}
\pdfoutput=1
\usepackage{jcappub}
\usepackage{graphicx, epsfig}
\usepackage{color}
\usepackage{mathrsfs}
\usepackage{bm}
\usepackage{amsmath,amssymb}
\usepackage[caption=false]{subfig}
\usepackage{hyperref}
\usepackage[normalem]{ulem}
\usepackage{mathtools}

\newcommand{\be}{\begin{equation}}
\newcommand{\ee}{\end{equation}}
\newcommand{\ba}{\begin{eqnarray}}
\newcommand{\ea}{\end{eqnarray}}

\begin{document}
\title{Classical-Quantum Correspondence and Hawking Radiation}
\author[a]{Tanmay Vachaspati,}
\author[a,b]{George Zahariade}
\affiliation[a]{Physics Department, Arizona State University, Tempe, AZ 85287, USA.}
\affiliation[b]{Beyond Center for Fundamental Concepts in Science, Arizona State University, Tempe, AZ 85287, USA.}

\abstract{
A recently discovered classical-quantum correspondence (CQC) maps certain quantum
problems to corresponding classical problems. We illustrate the CQC for a quantum scalar 
field in the gravitational background of a collapsing spherical shell. By truncating the problem 
to a small set of variables, we show how the CQC can yield Hawking radiation as well as the 
slow down of the collapse due to radiation backreaction.
}

\maketitle

\section{Introduction}

There are several time-dependent systems of interest that radiate quantumly
but not classically. Examples include particle production during inflation~\cite{Kofman:1994rk}, radiation
during gravitational collapse~\cite{Hawking:1974sw}, and systems of simple harmonic oscillators with quadratic
couplings~\cite{Vachaspati:2017jtw}. The basic idea is that the radiation field, $\phi$, in a specified time-dependent
background obeys a classical equation of the type,
\be
\square \Phi + m^2(t) \Phi = 0 
\label{waveeq}
\ee
where the D'Alembertian operator may include a time-dependent metric and $m^2$ may
arise due to interactions with a background dynamical field (such as the inflaton). If the
classical field $\Phi$ is taken to have trivial initial conditions, $\Phi (t=0,{\bf x}) =0={\dot \Phi} (t=0,{\bf x})$,
then Eq.~(\ref{waveeq}) is trivially solved as $\Phi (t,{\bf x})=0$ and there is no radiation. However,
quantumly, modes of the $\Phi$ field will get excited and there will be radiation. Such is the origin
of particle production at the end of inflation and Hawking radiation during gravitational collapse.
In this paper, we resolve this striking difference between classical and quantum evolutions by
finding non-trivial classical initial conditions such that the classical radiation rate is equal
to the quantum radiation rate \cite{Vachaspati:2018llo,Vachaspati:2018hcu}. Our analysis is quite general but will be 
concretely placed in the context of gravitational collapse, where the classical radiation is a classical
version of quantum Hawking radiation. 

Our result paves a way to study the backreaction of radiation on the background dynamics,
and is especially relevant to Hawking evaporation during gravitational collapse, a problem 
that has received
extensive attention in the past. This elusive phenomenon has been studied for instance in the context of the $1+1$ dimensional CGHS model~\cite{Callan:1992rs, Russo:1992ht, Susskind:1992gd} where the existence of stable remnants was discussed as a possible solution to Hawking's information paradox. Other authors have tried to approach the problem from a tunneling process point of view~\cite{Parikh:1999mf,Medved:2005yf}, or by considering the modified collapse of dust stars~\cite{Mersini-Houghton:2014zka, Mersini-Houghton:2014cta} or dust shells~\cite{Baccetti:2016lsb} due to emission of Hawking radiation (sometimes going so far as to suggest that horizon formation might not be the inevitable final state of gravitational collapse).
 
With suitable initial conditions in our approach, the classical energy emitted equals the quantum emission 
and so the classical and quantum evaporation rates are equal. Thus one may study Hawking evaporation
by solving classical equations of motion
in a related model and with specific initial conditions.
 The classical problem,
while conceptually easy, is still quite difficult technically. Hence we will instead reduce the problem further
to a toy minisuperspace problem for this initial investigation.

We start out in Sec.~\ref{cqc} by describing the classical-quantum correspondence
 \cite{Vachaspati:2018llo,Vachaspati:2018hcu}. We then describe the collapse of a spherical shell. Quantum fields in this
time-dependent background will be excited and the collapse will populate the radiative modes 
leading to Hawking radiation. This problem is too complicated to treat in full generality and so
we reduce it to a toy model with a few variables in Sec.~\ref{toymodel}. This truncation allows 
us to solve the system of classical equations and obtain the backreaction of the radiation on
the collapse. We discuss our findings in Sec.~\ref{discussion}.

\label{toymodel}

\section{The classical-quantum correspondence}
\label{cqc}

The classical-quantum correspondence (CQC) has been developed for quantum mechanical systems in Ref.~\cite{Vachaspati:2018llo} and for
quantum fields in Ref.~\cite{Vachaspati:2018hcu}. We will summarize those results here.

\subsection{The case of particles}
\label{cqcparticles}

In Ref.~\cite{Vachaspati:2018llo} the CQC was derived for quantum mechanical systems using the
Heisenberg picture. Here we will give the derivation in the Schrodinger picture.

Consider a simple harmonic oscillator (SHO), $a(t)$, with a time-dependent frequency $\omega (t)$.
The energy function is
\be
E = \frac{1}{2} {\dot{a}} ^2 + \frac{1}{2} \omega^2(t ) a^2
\ee
and the Schrodinger equation for the wavefunction $\psi (a,t)$ is,
\be
- \frac{1}{2} \frac{\partial^2 \psi }{\partial a^2} + \frac{1}{2} \omega^2 a^2 \psi 
= i \frac{\partial \psi }{\partial t} .
\ee
The lowest energy solution is~\cite{Lewis:1968yx}
\begin{equation}
\psi =  \frac{e^{i\gamma (t)}}{(2\pi \rho^2 ) ^{1/4}}
  \exp \left [ \frac{i}{2} \left ( \frac{\dot{\rho}}{\rho} + \frac{i}{2\rho^2} \right ) a^2 \right ]
\label{psisolution}
\end{equation}
where $\rho (t)$ is given by the real solution of the ordinary differential equation
\begin{equation}
\ddot{\rho} + \omega^2 (t) \rho = \frac{1}{4\rho^3}.
\label{rhoeq}
\end{equation}
Initial conditions for $\rho$ are chosen so that the wave function yields the ground
state of the time-independent SHO at $t=0$,
\begin{equation}
\rho (0) = \frac{1}{\sqrt{2\omega_0}}       \ , \ \ \ \dot{\rho}(0) =0
\label{rhoinitial}
\end{equation}
where $\omega_0$ is the frequency at the initial time. The phase $\gamma$ is given by $
\gamma (t ) = - \int_0^t  dt'/(4\rho^2 (t'))$. 
We can instead work in Cartesian coordinates by introducing two linearly 
independent classical degrees of freedom $\xi$ and $\chi$ such that $\rho=\sqrt{\xi^2+\chi^2}$ and that satisfy
\be
\ddot{\xi} + \omega^2 \xi =0, \ \  \ddot{\chi} + \omega^2 \chi =0
\ee
with the initial conditions
\be
\xi (0 ) = \frac{1}{\sqrt{2\omega_0}} , \ \
\dot{\xi} (0) = 0 , \ \ 
\chi (0) = 0 , \ \ 
\dot{\chi} (0) = \sqrt{\frac{\omega_0}{2}}\,.
\label{ic}
\ee
So we have now expressed the full dynamics of the quantum oscillator, namely the full time dependent
wavefunction in Eq.~\eqref{psisolution}, in terms of two classical oscillators given by $\xi$ and $\chi$.

From our explicit solution of the wavefunction we can calculate the energy that goes into mode 
excitations~\cite{Vachaspati:2006ki},
\be
E_q \equiv \langle\psi|H|\psi\rangle
= \frac{1}{2}\dot{\rho}^2+\frac{1}{2}\omega^2\rho^2+\frac{1}{8\rho^2}
\label{Eq}
\ee
where $H$ is the Hamiltonian operator. Note that this is exactly the polar coordinate expression for the classical energy of the {\it two-dimensional} 
harmonic oscillator with conserved angular momentum $1/2$ and initial energy $\omega_0/2$ whose dynamics 
are given by Eq. \eqref{rhoeq}.
After some algebra, the quantum energy in Eq.~(\ref{Eq}) can be written in terms of 
$\xi$ and $\chi$ as
\be
E_q = \frac{1}{2} {\dot{\xi}}^2+\frac{1}{2}\omega^2 \xi^2 
+  \frac{1}{2} {\dot{\chi}}^2+\frac{1}{2}\omega^2 \chi^2
\label{Eqsimple}
\ee
where we have used the invariance of the Wronskian, related to the angular
momentum of the simple harmonic oscillators,
\be
W \equiv \dot{\chi}\xi-\dot{\xi}\chi = \frac{1}{2}.
\label{wronskian}
\ee
Eq.~(\ref{Eqsimple}) is exactly the classical energy expression for {\it two} independent
simple harmonic oscillators, one given by the variable $\xi$ and the other by $\chi$, each
having initial conditions as in Eq.~(\ref{ic}) and each with initial energy $\omega_0/4$.
Further note that the relation holds for {\it any} time-dependence of $\omega(t)$.

Thus the quantum state of the SHO can be found by solving the classical system
in Eq.~(\ref{Eqsimple}) with initial conditions (\ref{ic}).
This correspondence is the key to 
simulating quantum radiation by using a classical calculation (also see Ref.~\cite{Vachaspati:2017jtw}). 
We expect it to be most useful 
in situations where the full quantum description is not available, such as in gravitational systems. 

\subsection{The case of fields}
\label{cqcfields}

Free quantum fields can be treated as an infinite set of SHOs.
For example, a real scalar field can always be expanded in a complete set of basis functions
(not necessarily eigenmodes),
\be
\phi(t,{\bf x})=\sum_k a_k(t)f_k({\bf x})\,,
\label{modedec}
\ee
and then the mode coefficients $a_k(t)$ are an infinite set of SHOs. Interaction with a 
space and time-dependent background will induce couplings between different mode
coefficients. If we write the action for $\phi$ as
\be
S_\phi = \int d^4x \sqrt{-g} \left [ \frac{1}{2} g^{\mu\nu} \partial_\mu\phi \partial_\nu\phi
- \frac{1}{2} m^2 \phi^2 \right ]
\ee
where the metric $g_{\mu\nu}$ and the mass function $m$ are arbitrary functions
of space and time, then the general Hamiltonian for the mode coefficients can be written 
as,
\be
H = \sum_{k,k'} \left [ \frac{1}{2} \pi_k {\bf N}_{kk'} \pi_{k'}  +  \frac{1}{2} a_k {\bf M}_{kk'} a_{k'} \right ]
\ee
where $\pi_k$ is the canonical momentum for the variable $a_k$, and ${\bf N}$ and ${\bf M}$ are 
time-dependent matrices. 

It has been shown in Ref.~\cite{Vachaspati:2018hcu} that this quantum problem for fields can also be mapped on to 
a classical problem.
If the modes are discretized and $N$ modes are considered, then the general classical 
problem involves $2N^2$ degrees of freedom. However, if a principal axis transformation
\cite{Kolopanis:2013sty} can be used to diagonalize ${\bf N}$ and ${\bf M}$ {\it for all times}, then the
CQC maps the quantum field $\phi$ to a complex classical scalar field that we call $\Phi$,
\be
S_\phi \to S_\Phi  = \int d^4x \sqrt{-g} \left [ \frac{1}{2} g^{\mu\nu} \partial_\mu\Phi^* \partial_\nu\Phi
- \frac{1}{2} m^2 \Phi^*\Phi \right ].
\label{complexcorr}
\ee
The quantum evolution of $\phi$ can then be written in terms of the classical evolution of $\Phi$
with specific initial conditions.

\subsection{Domain of applicability}

We should finally stress that the CQC will only be applicable to cases where the quantum degrees of freedom
can be, approximately or exactly, described as a set of quantum harmonic oscillators with time-dependent frequencies starting in their ground states, although the latter assumption can easily be dropped since considering am initial excited state only shifts the average energy by a constant; in other words the CQC encodes the effects of vacuum fluctuations over any state of the oscillator. This situation occurs in many interesting physical phenomena
involving a free radiation field in its vacuum state interacting with a classical time dependent background.

\section{A toy model for Hawking radiation backreaction}
\label{toymodel}

We are now in a position to discuss classical Hawking radiation. The
modes of a field in the spacetime background of a gravitationally collapsing object
get excited due to the time dependence of the collapse.
A quantum calculation of this ``fixed background'' process yields radiation that goes over 
to Hawking radiation in the infinite time limit~\cite{Vachaspati:2006ki,Kolopanis:2013sty}.
In Ref.~\cite{Vachaspati:2018llo} we have seen evidence that the CQC also yields the correct 
radiation backreaction on the background 
as long as the background can be treated classically. In the case discussed in
\cite{Vachaspati:2018llo} this meant that the dynamical time-scale had to be shorter
than the quantum time-scale associated with the spreading of the wavepacket. 
We expect similar criteria in the present case so that the CQC can be used to evaluate the
backreaction of Hawking radiation on the gravitational collapse for as long as the metric
can be treated classically. 
In Sec.~\ref{discussion} we will further discuss the timescale over which we expect
the CQC to hold.

To illustrate the backreaction of Hawking radiation, we consider a classical model based on
the gravitational collapse of a spherical vacuum shell with radius $R$. 
The full problem will require solution of Einstein's equations together with matter 
field equations but here we only consider a highly truncated toy model in which
we restrict ourselves to spherical geometries and assume coupling with one quantum
mode of a massless scalar field. 
We will see that, in a particular limit and as long as the radius of the shell can be treated
classically, this model reduces to a quantum harmonic oscillator initially in its ground state and with 
time-dependent frequency set by the dynamics of a classical background. Therefore the corresponding
backreaction problem can be treated via the CQC by replacing the quantum mode by two classical modes
with particular initial conditions.

The line element outside a vacuum shell of radius $R(t)$ is,
\begin{equation}
ds^2= -(1-\frac{R_S}{r}) dt^2 + (1-\frac{R_S}{r})^{-1} dr^2 + 
      r^2 d\Omega^2 \ ,
\label{metricexterior}
\end{equation}
where, $R_S = 2GM_{\rm shell}$ is the Schwarzschild radius in terms of the mass,
$M_{\rm shell}$, of the shell, and $d\Omega$ is the differential solid angle.
In the interior of the shell, the line element
is flat, as expected by Birkhoff's theorem, 
\begin{equation}
ds^2= -dT^2 +  dr^2 + r^2 d\Omega^2 \ .
\label{metricinterior}
\end{equation}
Following~\cite{Ipser:1983db} the mass of a collapsing shell in the absence of any 
other degrees of freedom is
\be
M_{\rm shell} = 4 \pi \sigma R^2 \left [ \frac{1}{\sqrt{1-R_T^2}} - 2\pi G\sigma R \right ]
\label{shellmass}
\ee
where $\sigma$ is the shell tension, $G$ is Newton's gravitational constant,
$R_T \equiv dR/dT$.
An action for the shell that leads to the conserved mass in Eq.~(\ref{shellmass}) is,
\be
S_{\rm shell} = - 4\pi \sigma \int dT R^2  \left [ \sqrt{1-R_T^2} - 2\pi G\sigma R \right ].
\label{Sshell}
\ee

In this metric, the action for a massless scalar field
can be split into regions inside the shell and outside the shell,
\ba
S_{\Phi} = 
-2 \pi \int dT \Biggl \{ \hskip -0.1 in &&\int_0^{R} dr ~ r^2 
 \left [
 - (\partial_T \Phi)^2 + (\partial_r \Phi)^2 
 \right ] 
\nonumber \\
+ 
&&\int_{R}^\infty dr ~ r^2
 \biggl [
 - \frac{{\dot T}(\partial_T \Phi)^2}{1-R_S/r} 
+ \frac{1}{\dot T} \left ( 1-\frac{R_S}{r} \right ) (\partial_r \Phi)^2 
  \biggr ] \Biggr \}
\label{actionterms2}
\ea
where 
\be
{\dot T} \equiv \frac{B}{\sqrt{B+(1-B)R_T^2}}
\label{dotT}
\ee
with 
\be
B \equiv 1- \frac{R_S}{R}\,.
\label{Bdefn}
\ee
For shell radius close to the Schwarzschild radius, $\dot{T}\approx -B/R_T\sim 0$, and the kinetic term is dominated by the integration 
within the shell while the gradient term is dominated by the integration outside the 
shell~\cite{Vachaspati:2006ki} so that the action reduces to
\be
S_\Phi\approx 2\pi\int dT\Biggl\{\int_0^{R_S}dr\,r^2(\partial_T\Phi)^2 -
\frac{1}{\dot T}\int_{R_S}^\infty dr\,r^2 \left ( 1-\frac{R_S}{r} \right) (\partial_r \Phi)^2\Biggl\}\ .
\label{SPhi}
\ee
Note that the time dependence of the metric leads to the overall factor of $1/{\dot T}$
in the gradient term.

Next we expand $\Phi$ in a {complete basis of function (not necessarily eigenmodes)} as in Eq. \eqref{modedec} 
and obtain a quadratic action for
the mode coefficients. 
Since the time dependence of the background enters as an overall factor of 
$1/{\dot T}$ in Eq.~(\ref{SPhi}),
a principal axis transformation~\cite{goldstein2002classical} 
will 
diagonalize this
action {\it for all times} and obtain the normal modes of the scalar field in (\ref{SPhi}). 
Each normal mode feels the changing 
metric of the shell due to the $1/{\dot T}$ coefficient of the second term.
Then the action for a normal mode coefficient is that of a quantum simple harmonic oscillator with a 
time-dependent frequency $\propto 1/\sqrt{\dot{T}}$. 

More explicitly, for a single mode of the field, we write $\Phi = z(t)f(r)$ and substitute in Eq.~\eqref{SPhi}.
Then we perform the radial integration and obtain a quadratic action for $z(t)$. The frequency of
the resulting simple harmonic oscillator is determined by the second integral in Eq.~\eqref{SPhi}.
In particular, the oscillator frequency cannot vanish except for the trivial shift mode for which
$\partial_r\Phi =0$.

Therefore, in this near-horizon approximation where different modes decouple, we explicitly truncate
the radiation field to one mode that describes a quantum harmonic oscillator with time-dependent frequency. 
Assuming it to be in its ground state at the onset of 
gravitational collapse, we can use the CQC and Eq.~(\ref{Eqsimple}) to
write the corresponding
action for two such classical simple harmonic oscillators,
\be
S_{\xi,\chi} = \int dT \left [ \frac{\xi_T^2}{2} - \frac{\kappa^2}{2{\dot T}} \xi^2
+\frac{\chi_T^2}{2} - \frac{\kappa^2}{2{\dot T}} \chi^2 \right ]
\label{Sxichi}
\ee
where a $T$ subscript denotes a derivative with respect to $T$ and
$\omega_0= \kappa/{(\dot T}(0))^{1/2}$.

A toy-model for backreaction of quantum radiation on the collapsing shell can thus be described by the 
equations of motion derived from the effective action $S_{\rm eff} = S_{\rm shell} + S_{\xi,\chi}$. 
However, as the collapsing shell couples to the classical simple harmonic oscillators $\xi$ and $\chi$, 
the mass $M_{\rm shell}$ appearing in the definition of the Schwarzschild radius needs to be replaced by
\be
M = R_T \pi_R - L_{\rm shell}
\label{Mdefn}
\ee
where $\pi_R = \partial L_{\rm eff} /\partial R_T$ is the momentum conjugate to $R$,
$L_{\rm eff}$ is the full Lagrangian and $L_{\rm shell}$ is the Lagrangian for the
shell as given by Eq.~(\ref{Sshell}). 

It is not possible to vary the action $S_{\rm eff}$ to obtain the exact equations of
motion because $M$ appears in ${\dot T}$ through $B$, and $M$ depends on $R$, 
$\xi$ and $\chi$ in a highly implicit manner. (Note that this difficulty arises only because we are
trying to find effective equations of motion for a much reduced system; the difficulty
will not arise in the full field theory discussed below.) We will therefore discuss two different approximations that will allow us to go forward.

\subsection{$M=M_{\rm shell}$ approximation}

Since the interaction terms in Eq. \eqref{Mdefn} are difficult to take into account, we 
simplify our analysis by ignoring the interaction energy in the mass. So we take,
\be
B \approx 1- \frac{2GM_{\rm shell}}{R} 
= 1- 2 \frac{R}{l} \left [ \frac{1}{\sqrt{1-R_T^2}} - \frac{R}{2l} \right ]
\label{Bdefn2}
\ee
with $l \equiv (4\pi G\sigma)^{-1}$. As we will see below, the 
interaction energy is negative and its magnitude increases with time. Therefore
the approximation in Eq.~(\ref{Bdefn2}) will eventually break down and a more refined approximation scheme will be required.

We note that $R$ only couples
to the combination $\xi^2 + \chi^2 = \rho^2$.
Using the conservation of the Wronskian and setting it to its initial value in Eq.~(\ref{wronskian}),
we can obtain the equations of motion for $R$ and $\rho$, which we write in terms of
rescaled variables $r=R/l$ (since there is no possible confusion with the previously used spherical radial coordinate), $\tau=T/l$ and $f = \rho /\sqrt{l}$,
\be
\frac{d}{d\tau} \left [ \frac{r^2 r_\tau}{\sqrt{1-r_\tau^2}}  \right ]
+ 2 r \sqrt{1-r_\tau^2} - \frac{3 r^2}{2} 
+ 
\frac{G \kappa^2}{2} \left [
\frac{d}{d\tau} 
\left ( f^2\frac{\partial }{\partial r_\tau} {\dot T}^{-1} \right )
- f^2 \frac{\partial}{\partial r} {\dot T}^{-1} \right ]
= 0\,,
\label{rdiffeq}
\ee
\begin{equation}
f_{\tau\tau} + (\kappa l)^2 {\dot T}^{-1} f = \frac{1}{4f^3}\,.
\label{feq}
\end{equation}
The initial conditions are
\be
r(0)= r_0, \ r_\tau(0)=0, \ f (0) = \frac{1}{\sqrt{2\omega_0 l}},\ f_\tau (0) =0\,.
\label{icfornumerics}
\ee
Note that we have $r_0 < 1$ because the shell is not initially a black hole~\cite{Ipser:1983db} and 
$r_\tau < 0$ because we are considering collapse. 

To solve the system of classical equations we need to specify the dimensionless
parameters: $r_0$, $G\kappa^2$, and $\kappa l$. We will consider,
\be
r_0 = 0.5, \ \ G\kappa^2 = 0.02, \ \ \kappa l = \frac{1}{r_0}
\ee
which give $\omega_0 l = 1/(r_0\sqrt{1-r_0})$. This corresponds to an initial shell mass of 
$13.29M_P$ where $M_P$ is the reduced Planck mass.

\begin{figure}[ht]
\begin{minipage}[t]{\dimexpr 0.5\textwidth-1em}
      \includegraphics[width=\textwidth,angle=0]{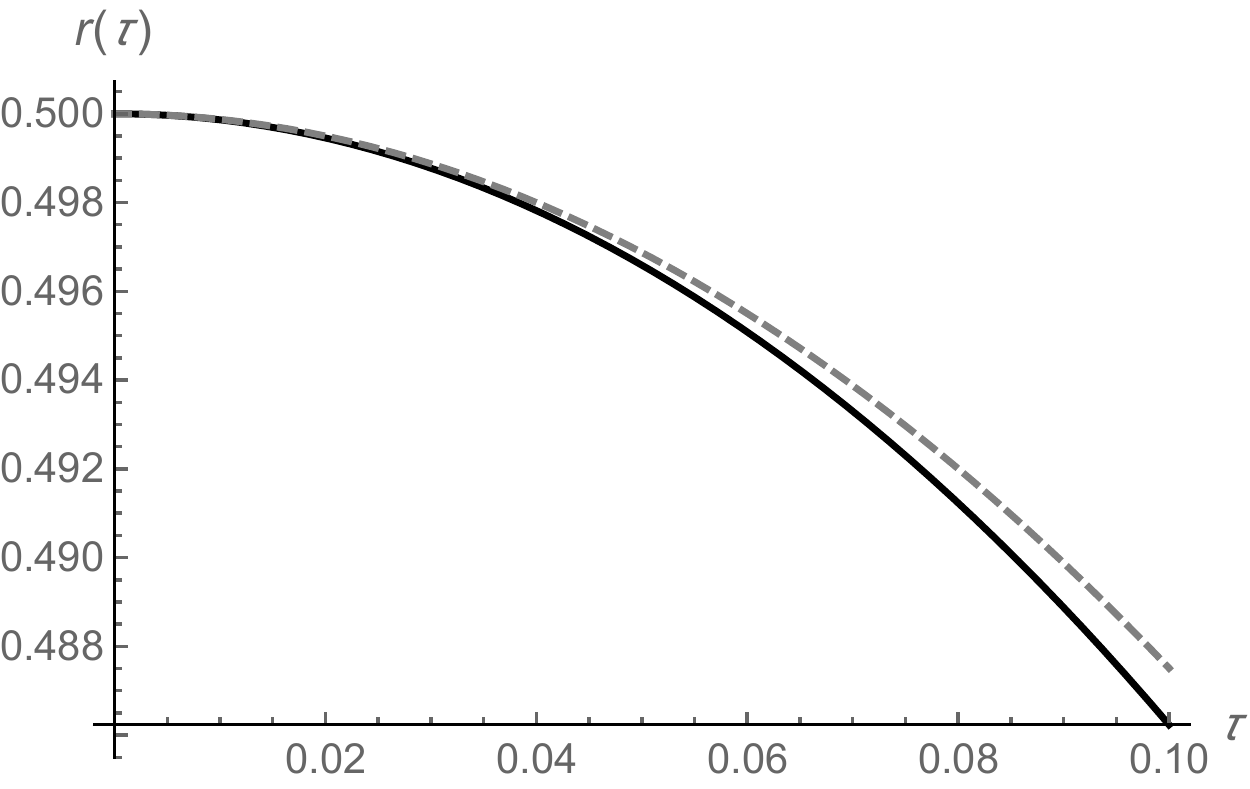}
  \caption{Radius of shell $r$ versus time $\tau$ when backreaction
  is ignored (dashed curve) and with backreaction taken into account (solid curve).
}
\label{radiustime}
\end{minipage}\hfill
\begin{minipage}[t]{\dimexpr 0.5\textwidth-1em}
      \includegraphics[width=\textwidth,angle=0]{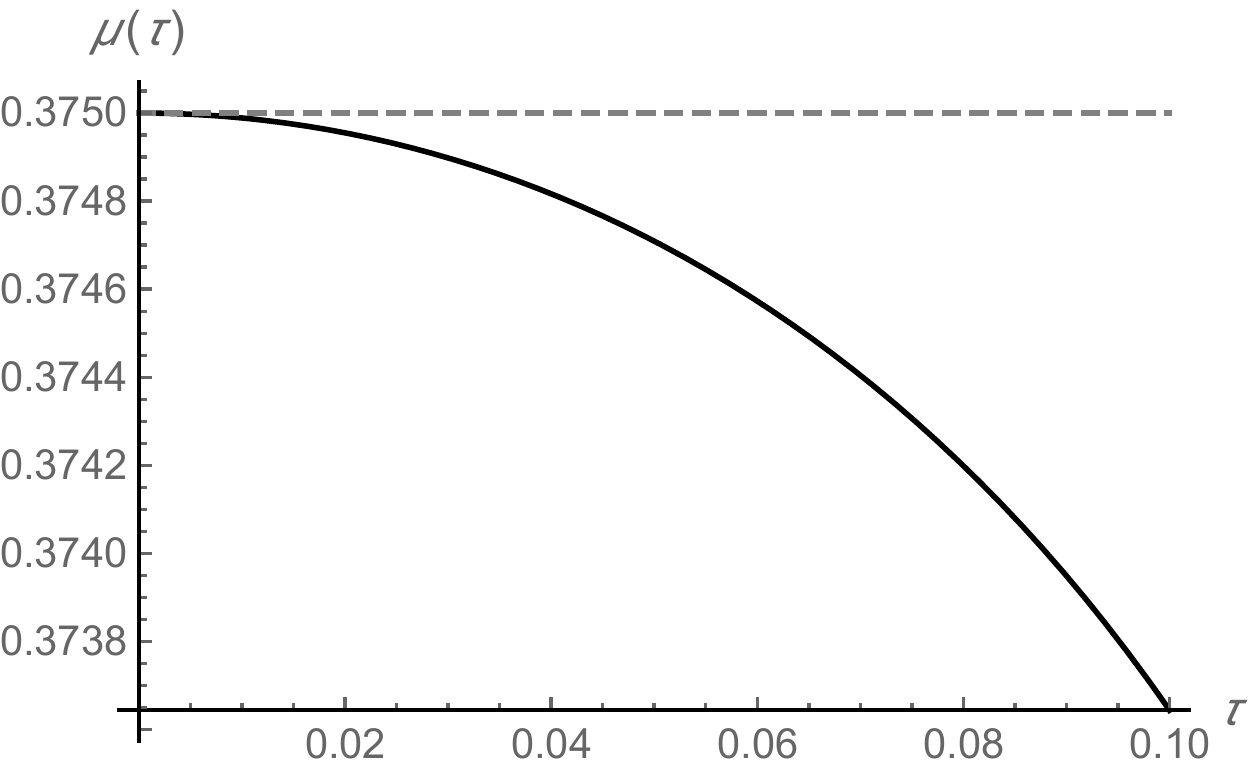}
  \caption{Rescaled Schwarzschild radius of the shell $\mu$ versus time $\tau$
 with (solid curve) and without (dashed curve) backreaction.}
\label{Mvstau}
\end{minipage}
\centering
\begin{minipage}[t]{\dimexpr 0.5\textwidth-1em}
  \includegraphics[width=\textwidth,angle=0]{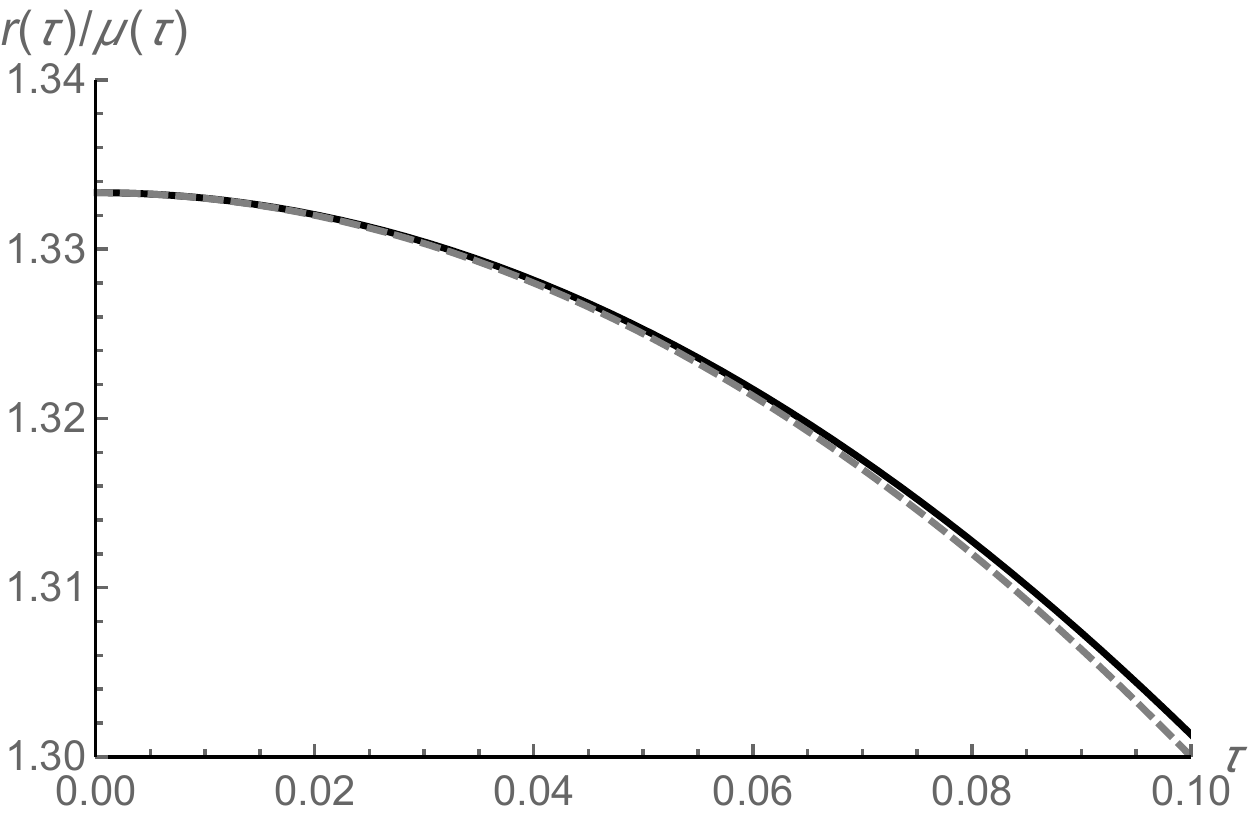}
  \caption{
  Radius of shell $r$ in units of the instantaneous (rescaled) Schwarzschild radius,
$\mu(\tau)$, versus time $\tau$ with (solid curve) and without (dashed curve) backreaction.
}
\label{rbyrstime}
\end{minipage}
\end{figure}

The numerical solutions for $r(\tau)$ with and without backreaction are shown in 
Fig.~\ref{radiustime}.
In the absence of backreaction, the collapse goes all the
way to $r=0$; with backreaction the collapse  
proceeds
faster but numerical
integration fails at a relatively early time, $\tau\approx0.1$. This is as we approach
the singularity in $1/{\dot T}$. In Fig.~\ref{Mvstau} we plot the rescaled Schwarzschild radius
\be
\mu\equiv \frac{2GM}{l}\approx 2r^2 \left [ \frac{1}{\sqrt{1-r_\tau^2}} - \frac{r}{2} \right ]\,,
\ee
(see Eq.~\eqref{Bdefn2}) as a function of time with and without backreaction,
showing how the shell evaporates. 

We have also checked that the total mass,
which includes the energy in the $\xi$ and $\chi$ fields in the case with
backreaction, is very well conserved, giving us confidence in
the numerical solution within this time interval. In Fig.~\ref{rbyrstime} we
plot the ratio of the shell radius to the instantaneous Schwarzschild
radius versus time. Here we see that backreaction causes the shell to
slow down relative to the decreasing Schwarzschild radius.

\subsection{$M$ determined via explicit energy conservation}

\begin{figure}[t]
\begin{minipage}[t]{\dimexpr 0.5\textwidth-1em}
      \includegraphics[width=\textwidth,angle=0]{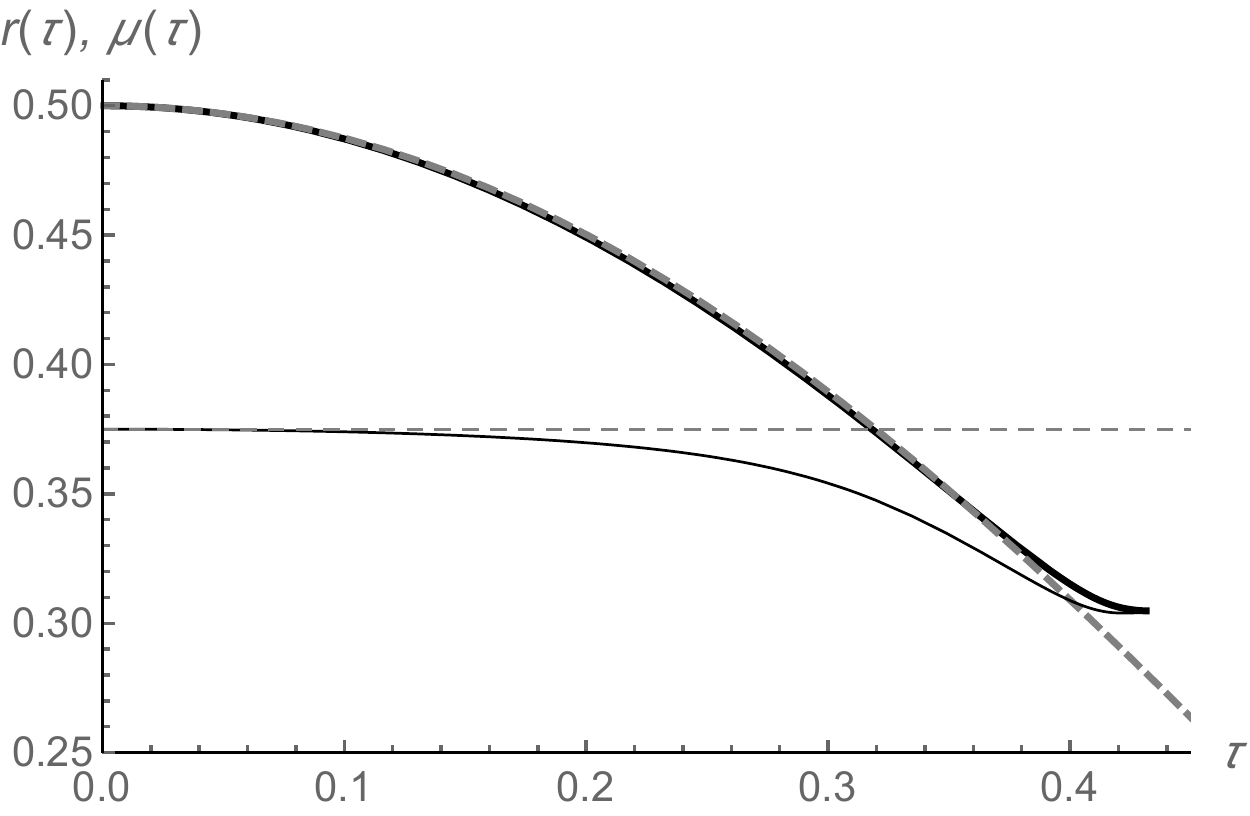}
  \caption{
  Rescaled radius $r$ of the shell (thick curves) and corresponding Schwarzschild radius 
  $\mu$ (thin curves) versus time $\tau$ when backreaction
  is ignored (dashed curves) and respectively with backreaction taken into account (solid curves).
}
\label{radiustimebis}
\end{minipage}\hfill
\begin{minipage}[t]{\dimexpr 0.5\textwidth-1em}
      \includegraphics[width=\textwidth,angle=0]{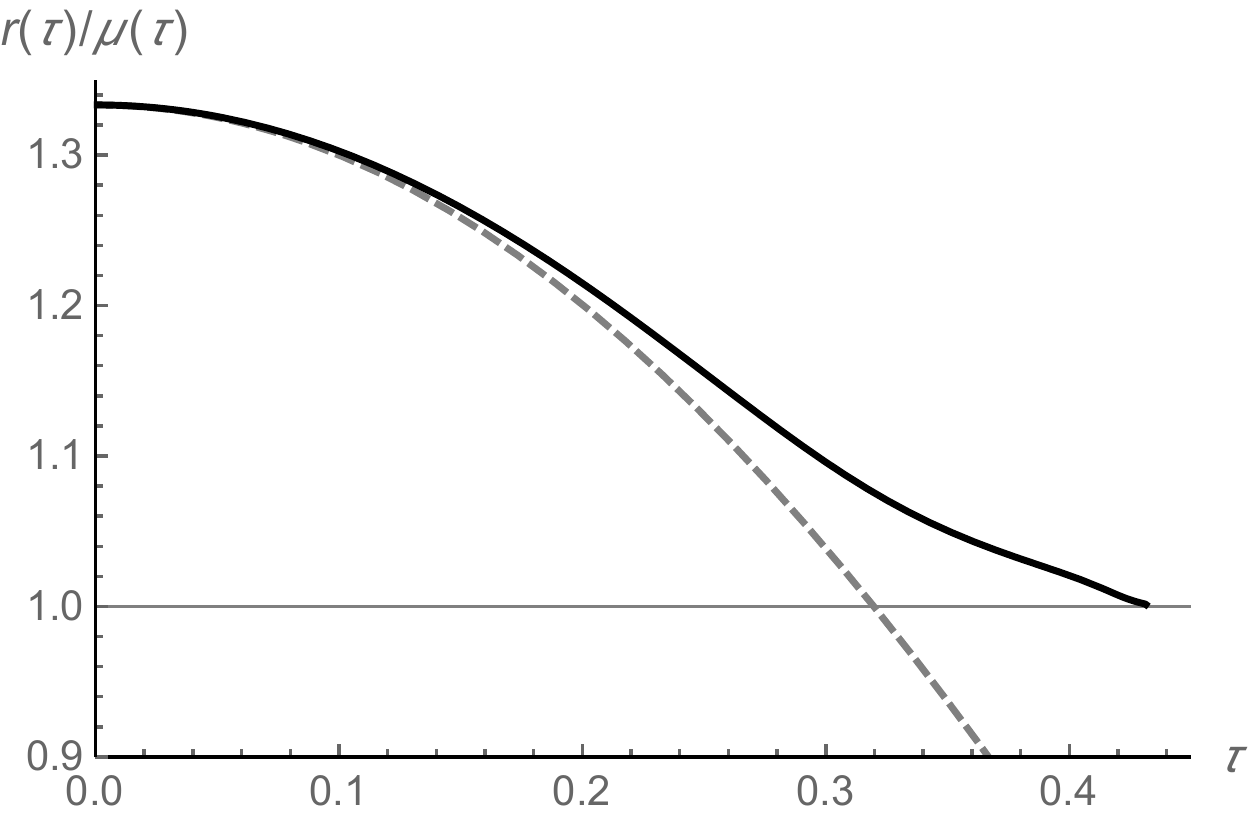}
  \caption{
  Radius of shell $r$ in units of the instantaneous (rescaled) Schwarzschild radius,
$\mu$, versus time $\tau$ with (solid curve) and without (dashed curve) backreaction.
}
\label{rbyrstimebis}
\end{minipage}
\end{figure}

In order to be able to carry out the numerical integration further, we need a better approximation for the mass $M$ of the collapsing wall. Since the evolution of
the full system conserves energy, we can use energy conservation to relate the 
rate of change of $M$ to the radiation rate.
In other words, to numerically solve the equations 
of motion we take $M(T)$ to be an additional degree of freedom and supplement 
the differential system describing the collapse dynamics with an additional constraint 
equation enforcing energy conservation,
\be
\frac{dM}{dT}=-\frac{d}{dT}\left [ \frac{\xi_T^2}{2} + \frac{\kappa^2}{2{\dot T}} \xi^2
+\frac{\chi_T^2}{2} + \frac{\kappa^2}{2{\dot T}} \chi^2 \right ]\ .
\label{masstimevar}
\ee
This equation simply states that all the energy that is radiated into the modes $\xi$ and $\chi$ must 
come from the shell itself {\it i.e.} from $M(T)$.
Therefore the dynamics of the collapse are described by
the equations of motion for $R$, $\xi$ and $\chi$ (with $M(T)$ considered to be a spectator function) 
as well as Eq.~\eqref{masstimevar}. In terms of the rescaled functions $r(\tau)$, $f(\tau)$ and $\mu(\tau)$, the relevant differential equations will be Eqs. \eqref{rdiffeq} and \eqref{feq} (which now depend explicitly on $\mu(\tau)$ via $\dot{T}$) as well as
\be
\frac{d\mu}{d\tau}+G\kappa^2 f^2\frac{d}{d\tau}\left(\dot{T}^{-1}\right)=0\,.
\label{mueq}
\ee
Note that Eq. \eqref{mueq} is an implicit equation for $d\mu/d\tau$ (since the second
term also involves $d\mu/d\tau$)
and we have used Eq. \eqref{feq} to simplify its expression.
The initial conditions are those given in Eq. \eqref{icfornumerics} along with
\be
\mu(0)= 2r_0^2\left[1-r_0/2 \right]\,.
\ee
The system of classical equations is solved for the same choice of parameters as in the previous section.
The numerical solutions for $r(\tau)$ and $\mu(\tau)$ with and without backreaction are shown in 
Fig.~\ref{radiustimebis}.
In the absence of backreaction, the collapse would go all the
way to $r=0$ and the rescaled Schwarzschild radius $\mu$ would remain constant. 
With backreaction the shell evaporates (as illustrated by the monotonic decrease of $\mu$) 
and the collapse initially proceeds at approximately the same rate as in the non-radiating case.
However the evaporation is then seen to slow down as both $r$ and $\mu$ seem to stabilize 
around the same non-zero value  before numerical
integration fails. This is as we approach
the singularity in $1/{\dot T}$.
To make this observation more precise, in Fig.~\ref{rbyrstimebis} we
plot the ratio of the shell radius to the instantaneous Schwarzschild
radius versus time. Here too we see that backreaction causes the shell collapse to
slow down relative to the decreasing Schwarzschild radius while the ratio $r/\mu$ seems to asymptote to one.

This new approximation scheme therefore allows us to trace the evaporation of the shell on a longer time scale. We can thus see that the behavior encountered in the previous subsection with the collapse initially proceeding faster was just a small transient feature not representative of the asymptotic dynamics of the shell. 

\section{Discussion}
\label{discussion}

Despite these intriguing preliminary results we do not expect our effective
model to remain a good description as the evaporation proceeds 
and interactions become dominant, 
since the assumptions allowing us to treat the shell classically and 
use the CQC may not hold anymore, and different radiative modes may become coupled to each other.
An alternative
is to start with a full field theory description and to solve the classical field
equations with appropriate boundary conditions. For example, as discussed in Sec. \ref{cqcfields} and in Ref. \cite{Vachaspati:2018hcu}, in the particular case where a principal axis diagonalization can be done independently of time, a field theory 
action which has all the necessary ingredients is 
\be
S = S_{EH} + S_{\lambda\phi^4} + S_{\psi_1,\psi_2}
\label{fullS}
\ee
where the first term is the Einstein-Hilbert action, the second is the action for a
scalar field, $\phi$, with a double well potential which is known to have domain wall solutions,
and the last term is the action for two massless scalar fields $\psi_1$ and $\psi_2$. The
initial conditions for $\phi$ will have to be chosen to represent a collapsing spherical
domain wall, and for $\psi_1$ and $\psi_2$ the initial conditions will have to be chosen 
according to Eq.~(\ref{ic}). 
Note that $\psi_1$ and $\psi_2$ are equivalent to
a single complex scalar field with a global U(1) symmetry 
(see Eq. \eqref{complexcorr}). The non-zero Wronskian for the simple 
harmonic oscillators in Eq.~(\ref{wronskian}) is equivalent to having a non-zero global U(1) 
charge in the system that, like the Wronskian, is conserved during collapse. For the
simple harmonic oscillator system we have discussed, the conservation of the Wronskian
has important dynamical implications since it provides a centrifugal barrier for the
dynamics. It seems likely that the conservation of the global U(1) charge will also have
important consequences in the full field theory.

To summarize, a key gap in our understanding of gravitational collapse is the backreaction
of quantum radiation on the collapse. This is difficult because a proper
treatment requires quantization of the spacetime metric while we do not yet 
have an adequate quantum theory of gravity. Our results show
that there is an analogous classical backreaction problem in which the radiation
and backreaction (evaporation) can be evaluated by solving classical 
equations of motion. This method is valid as long as the radiation can be modeled (at least in some limit) as a free field
in its vacuum state living in a time-dependent classical background. In a simplified model we show that the collapse
proceeds slower in units of the Schwarzschild radius when backreaction is taken into 
account and that it even shows signs of plateauing before a black hole is formed (a peculiar feature also hinted at in~\cite{Baccetti:2016lsb}), but the model 
breaks down by the time of horizon formation. 

More precisely, we start out with a macroscopic collapsing object
that we assume can be accurately described as classical. As the collapse proceeds, the object 
loses energy to the radiation field and its degrees of freedom become entangled with the 
radiation modes, so that the classical description will inevitably break down at some time. 
An estimate of the duration of validity of the CQC is obtained by noting that 
a full quantum treatment of the shell will become necessary when the
occupation number of the radiation field becomes comparable to the occupation number
for the collapsing shell: $N_r \sim N_s$. The occupation number of the shell may be
estimated as the energy of the shell divided by the energy scale of the shell, $\sigma^{1/3}$,
where $\sigma$ is the shell tension.
The occupation number of the radiation is the energy in the radiation (which increases
with time) divided by the
typical frequency, which is given by the Hawking temperature, $T_H=(8\pi G M)^{-1}$. 
Then an estimate for the duration of validity of the CQC is given by
\be
\frac{M}{\sigma^{1/3}} \sim \frac{T_H^2 t_{\rm CQC}}{T_H}\quad\text{or}\quad 
t_{\rm CQC} \sim \frac{8\pi GM^2}{\sigma^{1/3}}\,,
\ee
which is much longer than the numerical integration time. We have also checked that 
our numerical analysis is only for times within the
duration of validity of the CQC by estimating the occupation number of the radiation
mode directly via the relation $N_r\sim(M-M_0)/\omega_0$.

Our work distinguishes itself from previous studies in a few important ways. First of all, it does 
not require anything more than standard four-dimensional GR and the theory of quantum fields 
in curved spacetime. Indeed a lot of interesting work has been done in lower dimensional models 
(sometimes motivated by string theory~\cite{Callan:1992rs, Russo:1992ht, Susskind:1992gd}). 
Second, the radiation is treated fully quantumly and the computations are carried out in a classical 
system with non-trivial initial conditions which \emph{exactly} simulates the quantum fluctuations of 
the radiation field. It is important to insist on the fact that nowhere do we rely on an effective classical 
description of quantum evaporation during gravitational collapse~\cite{Mersini-Houghton:2014zka, Mersini-Houghton:2014cta,Baccetti:2016lsb}. Nor are we required to interpret the horizon as a potential barrier through 
which particles can tunnel~\cite{Parikh:1999mf,Medved:2005yf}. Finally, despite the approximations 
used in order to have tractable numerical computations our approach is quite conservative and can 
in principle be used to study more realistic models of gravitational collapse in the presence of quantum 
radiation. The two simplified models of a collapsing spherically symmetric domain wall showcased in this 
paper provide a powerful proof of principle.
We expect an analysis in a more complete field theory model (Eq.~(\ref{fullS})) can
provide insight into the fate of gravitational collapse and the formation and nature of black 
holes.

\acknowledgments
We are grateful for comments by David Garfinkle and Juan Maldacena.
TV's work is supported by the U.S. Department of Energy, 
Office of High Energy Physics, under Award No. DE-SC0018330 at Arizona State University
and GZ is supported by John Templeton Foundation grant 60253. GZ would like to thank the 
Yukawa Institute for Theoretical Physics as well as Tohoku University for support during the 
completion of this work.

\bibliographystyle{JHEP.bst}
\bibliography{classicalHawking}

\providecommand{\href}[2]{#2}\begingroup\raggedright\begin{thebibliography}{10}

\bibitem{Kofman:1994rk}
L.~Kofman, A.~D. Linde and A.~A. Starobinsky, \emph{{Reheating after
  inflation}}, \href{https://doi.org/10.1103/PhysRevLett.73.3195}{\emph{Phys.
  Rev. Lett.} {\bfseries 73} (1994) 3195}
  [\href{https://arxiv.org/abs/hep-th/9405187}{{\ttfamily hep-th/9405187}}].

\bibitem{Hawking:1974sw}
S.~W. Hawking, \emph{{Particle Creation by Black Holes}},
  \href{https://doi.org/10.1007/BF02345020}{\emph{Commun. Math. Phys.}
  {\bfseries 43} (1975) 199}.

\bibitem{Vachaspati:2017jtw}
T.~Vachaspati, \emph{{Quantum Backreaction on Classical Dynamics}},
  \href{https://doi.org/10.1103/PhysRevD.95.125002}{\emph{Phys. Rev.}
  {\bfseries D95} (2017) 125002}
  [\href{https://arxiv.org/abs/1704.06235}{{\ttfamily 1704.06235}}].

\bibitem{Vachaspati:2018llo}
T.~Vachaspati and G.~Zahariade, \emph{{A Classical-Quantum Correspondence and
  Backreaction}},  \href{https://arxiv.org/abs/1806.05196}{{\ttfamily
  1806.05196}}.

\bibitem{Vachaspati:2018hcu}
T.~Vachaspati and G.~Zahariade, \emph{{Classical-Quantum Correspondence for
  Fields}},  \href{https://arxiv.org/abs/1807.10282}{{\ttfamily 1807.10282}}.

\bibitem{Callan:1992rs}
C.~G. Callan, Jr., S.~B. Giddings, J.~A. Harvey and A.~Strominger,
  \emph{{Evanescent black holes}},
  \href{https://doi.org/10.1103/PhysRevD.45.R1005}{\emph{Phys. Rev.} {\bfseries
  D45} (1992) R1005} [\href{https://arxiv.org/abs/hep-th/9111056}{{\ttfamily
  hep-th/9111056}}].

\bibitem{Russo:1992ht}
J.~G. Russo, L.~Susskind and L.~Thorlacius, \emph{{Black hole evaporation in
  (1+1)-dimensions}},
  \href{https://doi.org/10.1016/0370-2693(92)90601-Y}{\emph{Phys. Lett.}
  {\bfseries B292} (1992) 13}
  [\href{https://arxiv.org/abs/hep-th/9201074}{{\ttfamily hep-th/9201074}}].

\bibitem{Susskind:1992gd}
L.~Susskind and L.~Thorlacius, \emph{{Hawking radiation and back reaction}},
  \href{https://doi.org/10.1016/0550-3213(92)90081-L}{\emph{Nucl. Phys.}
  {\bfseries B382} (1992) 123}
  [\href{https://arxiv.org/abs/hep-th/9203054}{{\ttfamily hep-th/9203054}}].

\bibitem{Parikh:1999mf}
M.~K. Parikh and F.~Wilczek, \emph{{Hawking radiation as tunneling}},
  \href{https://doi.org/10.1103/PhysRevLett.85.5042}{\emph{Phys. Rev. Lett.}
  {\bfseries 85} (2000) 5042}
  [\href{https://arxiv.org/abs/hep-th/9907001}{{\ttfamily hep-th/9907001}}].

\bibitem{Medved:2005yf}
A.~J.~M. Medved and E.~C. Vagenas, \emph{{On Hawking radiation as tunneling
  with back-reaction}},
  \href{https://doi.org/10.1142/S021773230501861X}{\emph{Mod. Phys. Lett.}
  {\bfseries A20} (2005) 2449}
  [\href{https://arxiv.org/abs/gr-qc/0504113}{{\ttfamily gr-qc/0504113}}].

\bibitem{Mersini-Houghton:2014zka}
L.~Mersini-Houghton, \emph{{Backreaction of Hawking Radiation on a
  Gravitationally Collapsing Star I: Black Holes?}},
  \href{https://doi.org/10.1016/j.physletb.2014.09.018}{\emph{Phys. Lett.}
  {\bfseries B738} (2014) 61}
  [\href{https://arxiv.org/abs/1406.1525}{{\ttfamily 1406.1525}}].

\bibitem{Mersini-Houghton:2014cta}
L.~Mersini-Houghton and H.~P. Pfeiffer, \emph{{Back-reaction of the Hawking
  radiation flux on a gravitationally collapsing star II}},
  \href{https://arxiv.org/abs/1409.1837}{{\ttfamily 1409.1837}}.

\bibitem{Baccetti:2016lsb}
V.~Baccetti, R.~B. Mann and D.~R. Terno, \emph{{Role of evaporation in
  gravitational collapse}},
  \href{https://doi.org/10.1088/1361-6382/aad70e}{\emph{Class. Quant. Grav.}
  {\bfseries 35} (2018) 185005}
  [\href{https://arxiv.org/abs/1610.07839}{{\ttfamily 1610.07839}}].

\bibitem{Lewis:1968yx}
H.~R. Lewis, \emph{{Class of exact invariants for classical and quantum
  time-dependent harmonic oscillators}},
  \href{https://doi.org/10.1063/1.1664532}{\emph{J. Math. Phys.} {\bfseries 9}
  (1968) 1976}.

\bibitem{Vachaspati:2006ki}
T.~Vachaspati, D.~Stojkovic and L.~M. Krauss, \emph{{Observation of incipient
  black holes and the information loss problem}},
  \href{https://doi.org/10.1103/PhysRevD.76.024005}{\emph{Phys. Rev.}
  {\bfseries D76} (2007) 024005}
  [\href{https://arxiv.org/abs/gr-qc/0609024}{{\ttfamily gr-qc/0609024}}].

\bibitem{Kolopanis:2013sty}
M.~Kolopanis and T.~Vachaspati, \emph{{Quantum Excitations in Time-Dependent
  Backgrounds}}, \href{https://doi.org/10.1103/PhysRevD.87.085041}{\emph{Phys.
  Rev.} {\bfseries D87} (2013) 085041}
  [\href{https://arxiv.org/abs/1302.1449}{{\ttfamily 1302.1449}}].

\bibitem{Ipser:1983db}
J.~Ipser and P.~Sikivie, \emph{{The Gravitationally Repulsive Domain Wall}},
  \href{https://doi.org/10.1103/PhysRevD.30.712}{\emph{Phys. Rev.} {\bfseries
  D30} (1984) 712}.

\bibitem{goldstein2002classical}
H.~Goldstein, C.~Poole and J.~Safko, \emph{Classical Mechanics}. Addison
  Wesley, 2002.

\end{thebibliography}\endgroup

\end{document}